# First analysis of results of the OH survey of the Inner Galaxy: evidence for a stellar bar

M.N. Sevenster

*Sterrewacht Leiden, P.O.Box 9513, 2300 RA Leiden, NL*

**Abstract.** Part of a large survey of the inner galactic Plane ( $|\ell| < 45°$ and $|b| < 3°$ total) in the OH 1612MHz line in search for OH/IR stars is analysed. We find strong evidence for a central m=2 distortion based on geometrical considerations. The observed deviation from axisymmetry cannot be explained by lopsidedness and agrees with other recent models of the galactic Bar on length, inclination and axis ratio.

## 1. Introduction

OH/IR stars are far-evolved objects at the end of Asymptotic Giant Branch (AGB) evolution. The central star is totally obscured by a circumstellar envelope (CSE) that emits strong maser emission in the OH line at 18 cm. The CSE expands because of radiation pressure on the dust it contains. Hence we see a red- and a blue-shifted component in the spectrum that we can average to find the stellar velocity. The expansion velocity increases with stellar luminosity and with CSE metallicity.

In this paper we discuss a sample of such stars. One of the goals of the survey that created this sample was to find evidence for a bar in the Milky Way Galaxy. There is increasing evidence for non-axisymmetry of the Galaxy (see Kuijken, these proceedings). It has so far been difficult to eliminate the possibility of the distortion being a lopsidedness or m=1 distortion, which is not totally unlikely in view of observations of other galaxies. Stellar kinematics (Zhao *et al.* 1994) or microlensing events (OGLE, Stanek 1995) are most suited to find evidence for an m=2 distortion or bar, but for such analysis it is difficult to acquire a large enough sample.

The present survey has a homogeneous and extended spatial coverage, both in the sky and along the line of sight, and accurate radial velocities. In this paper we present evidence for a bar in this paper based purely on simple morphological arguments. The radial velocities are not yet discussed.

## 2. Observations

The observations discussed in this paper were taken with the Australia Telescope Compact Array. The region between $|\ell| < 10°$ and $|b| < 3°$ was completely covered, with pointings separated by one half power beamwidth of the primary beam. The bandwidth of the observations covered velocities between -320 $\mathrm{km\,s^{-1}}$ and +390 $\mathrm{km\,s^{-1}}$ which is sufficient to find all stars except a

few extreme velocity outliers. This resulted in a set of 317 stars, 242 of which have well-determined stellar velocities with an accuracy of 1.5 km s$^{-1}$, with 1″ spatial resolution.

## 3. A galactic Bar in projection

Any set of observations will have a certain distance limit. In this section the appearance of a non-axisymmetry in the number of stars along the line-of-sight as a function of galactic longitude N($\ell$) for samples is discussed. The form of N($\ell$) depends on the distance $d$ out to which the Galaxy is sampled by observations. We calculate N($\ell,d$) for various ($\ell,d$) for a two-dimensional elliptical bar with gaussian density distribution. The results are shown in Fig. 1.

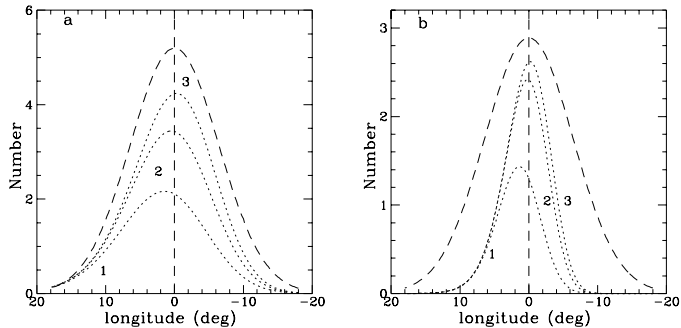

**Fig. 1** *Total number of stars along the line-of-sight in a two-dimensional elliptical barmodel with gaussian density distribution. Integration limits are 8 kpc (1), 9 kpc (2) and absent (3). The bar is inclined with respect to the line-of-sight to the GC by 20° with the near end at positive longitudes. For reference a gaussian is shown (dashed). The distance to the GC is taken to be 8 kpc. a Axis ratio 0.6, semi major axis 3.5 kpc. b Axis ratio 0.4, semi major axis 2.5 kpc.*

For the small integration limits N($\ell,d$) essentially looks like the distribution arising from an m=1 distortion at $\ell > 0°$, with its maximum toward *positive* longitudes. Without integration limit the distribution is skewed with its maximum toward *negative* longitudes for the same model m=2 distortion. This is the result of the line-of-sight through the m=2 distortion being longer on the far side of the distortion than the near side for small values of absolute longitudes.

The strength of this effect depends on axis ratio, major axis, density distribution and the inclination of the bar (the in-plane rotation angle).

## 4. A galactic Bar in the observations

The expansion velocity of the CSE is related to the intrinsic stellar luminosity and the CSE gas-to-dust ratio $\mu$, by :

$$\mu \propto L_*^{0.5} v_{exp}^{-2} \qquad \text{(van der Veen 1989)}$$

We divide the sample of doubly peaked OH/IR stars into two, with average $v_{exp}$ of 11.3 (sample I) respectively 18.3 (sample II) km s$^{-1}$. This gives a factor of 1.7 difference in stellar luminosity $L_*$, if we assume $\mu_I = 2\mu_{II}$ . (Blommaert (1992) found a range in $\mu$ of $\sim 2.5$ in the GC with IR observations. The range of $L_*$ in the Bulge is found to be $\sim 1 - 10 \times 10^4 L_\odot$ (van der Veen and Habing 1990) which agrees well with a factor of 1.7 between the samples.) Since the OH masers are saturated the OH luminosity, $L_{OH}$, increases linearly with $L_*$. The limiting flux $F_{OH}$ is naturally the same for both samples, so the average limiting distance of sample II is a factor $\sim 1.3$ larger than of sample I.

The effect of skewed distributions will be clearest in the inner regions of the Galaxy (Fig. 1). The ratios of the number of stars with $0° < \ell < 4°$ to the number of stars with $0° > \ell > -4°$ are 39/35 (sample I) and 22/35 (sample II). These ratios are in accordance with the results in Sec. 3. To better define these trends, we sorted both samples on their *absolute* longitudes and then calculate the cumulative sums of $\ell/|\ell|$: we add or substract 1 for each star. An axisymmetric distribution gives a line that hovers around zero. If negative (positive) longitudes are 'overpopulated' the sum will steadily decay (rise).

This relation is shown in Fig. 2 for the two data sets and for the bar model shown in Fig. 1b . This model is similar to those derived from the COBE data (Dwek *et al.* ) with an inclination of 20° , a length of 2.5 kpc, an axis ratio in the plane of 0.4 and no tilt out of the plane.

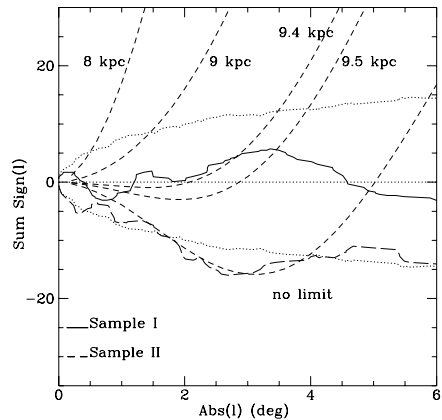

**Fig. 2** *The cumulative sum $\Sigma(\ell/|\ell|)$ versus $|\ell|$ after sorting on $|\ell|$ for the two samples (solid and dashed lines) and for a bar model with different integration cut-offs (short dashed lines). The dotted lines indicate the values for which the chance of the sum arising from a binomial distribution being bigger (smaller for negative values) than that value is 5%. It should be noted that for $|\ell| < 0.5°$ and $|\ell| > 5°$ contributions by GC and Disk or Ring stars cannot be ignored.*

Sample I (solid line) never deviates significantly from axisymmetry, although there are clear local trends. Sample II (dashed line) , however, lies at or outside the 5% confidence level (dotted lines), which indicates a significant deviation from axisymmetry. The model with the intermediate cut-off shows local trends similar to sample I, this may explain why mere comparison with a global binomial distribution is not enough to show a significant deviation. The model without distance cut-off coincides remarkably well with the sample II.

## 5. Conclusions

Comparing two subsets of a sample of OH/IR stars that differ in average distance from the Sun we find that their longitude distributions differ significantly. This can be explained by assuming that the inner Galaxy is barred and that the two sets sample the bar to different distances. The observations agree with models for the galactic Bar from COBE data with an inclination of 20°, axis ratio of 0.4 and a semi-major axis of 2.5 kpc. In our model, the set of stars with low $v_{\rm exp}$ most likely has a distance cut-off of around 9 kpc, the set with high $v_{\rm exp}$ of at least 11 kpc. This coincides with the difference of a factor 1.3 in average distance derived from the relation between $v_{\rm exp}$ and stellar luminosity. This is the first large scale morphological evidence for a galactic Bar that cannot also be explained by a lopsided Galaxy.

**Acknowledgments.** I thank Richard Arnold, Harm Habing and Frank van der Bosch for critical reading and listening.

### Discussion

*P. Teuben*: Can you from these data already exclude the dynamical centre being around 100pc from the galactic Centre towards positive longitudes ?

*M. Sevenster*: I cannot exclude an offset of the dynamical centre in any direction so far. (We cannot take the mean galactic longitude to be the longitude of the dynamical centre, because of exactly the projection effects discussed in this paper. In the used model ,however, 100pc is too much to reproduce the observed over-abundance of high $v_{\rm exp}$ stars at negative longitudes.)

*W. van Driel*: Unfortunately, your OH line data suffer from interference, varying from day to day. This can cause quite peculiar and disturbing biases in the source distribution. Do you intend to eliminate this bias by re-observing the data suffering from interference ?

*M. Sevenster*: No, there will be no re-observing. I intend to eliminate or at least diminish this effect by weighing the data according to their noiselevels.